\documentclass{icrc2009}

\usepackage{graphicx}   
\usepackage{caption}    
\usepackage{subfig} 
\usepackage{fixltx2e}
\usepackage{url}

\newcommand{\shorttitle}[1]%
{\markboth{Proceedings of the 31\MakeLowercase{$^{st}$} ICRC, {\L}\'{o}d\'{z} 2009}{#1} }
\newcommand{\etal}{\MakeLowercase{\textit{et al. }}} 

\begin{document}
\title{Data Analysis for the Measurement of High Energy Cosmic Ray Electron/Positron Spectrum with Fermi-LAT
}

\author{\IEEEauthorblockN{M.N.~Mazziotta\IEEEauthorrefmark{1} on behalf of the Fermi-LAT collaboration \\ \IEEEauthorblockA{\IEEEauthorrefmark{1} Istituto Nazionale di Fisica Nucleare-Sezione di Bari, via Orabona 4, I-70126 Bari, Italy}}}

\shorttitle{M.N.~Mazziotta \etal Data Analysis for High Energy Electron/Positron with Fermi-LAT}
\maketitle

\begin{abstract}
The Large Area Telescope (LAT) instrument on board the Fermi satellite consists of a multi-layer silicon-strip tracker interleaved with tungsten converters (TKR), followed by a CsI crystal hodoscopic calorimeter (CAL). Sixteen TKR and CAL modules are assembled in a 4$\times$4 array. A segmented anticoincidence plastic scintillator (ACD) surrounds the TKRs. The primary cosmic-ray electron/positron energy spectrum has been measured from 20 GeV to 1 TeV using a dedicated event analysis that ensures efficient electron detection and reduced hadron contamination. Results from detailed Monte Carlo simulations have been used to reconstruct the observed energy spectrum to the primary cosmic ray spectrum. We present here details of the analysis procedure and the energy spectrum reconstruction.
  \end{abstract}

 
\section{Introduction}
The electron component in the primary cosmic radiation (CR) consists of both electrons and positrons. A number of measurements of their energy spectrum by balloon-borne experiments \cite{1} and by a single space mission \cite{2} have shown that the electron intensity, about 1\% of the proton intensity at 10 GeV, decreases more rapidly with energy than that of cosmic-ray nuclei. Although existing data exhibit sizeable statistical and systematic uncertainties, it is clear that the power-law index of the electron spectrum above 10 GeV is larger than 3.0, in contrast to the proton index of 2.7.
High-energy electrons in the galactic magnetic field lose energy through synchrotron radiation and by inverse Compton collisions with low-energy photons. The measurement of the energy spectrum of primary electrons can give information on the galactic magnetic field as well as on the confinement time of electrons in the Galaxy and on the distribution of particle sources in our Galaxy. Positrons are believed to be mainly created in secondary production processes resulting from the interaction of CR nuclei with the interstellar medium. 

Early measurements \cite{3} capable of separating electrons and positrons have shown that the positron fraction $e^+/(e^+ + e^-)$ is around 10\% in the 1-100 GeV region, even though this data deviate significantly from predictions of secondary production models.

In this paper we report details of the analysis procedure and energy spectrum reconstruction using of data collected in the first six months of operation by the Large Area Telescope (LAT) instrument on board the Fermi satellite \cite{prl}.

\section{Detector description}
The LAT is a gamma-ray Telescope based on the conversion of gamma-rays into electron-positron pairs and is arranged in a $4 \times 4$ array of 16 identical towers. Each tower consists of a silicon micro-strip detectors Tracker (TKR) followed by a segmented CsI calorimeter (CAL), to reconstruct the gamma-ray direction and energy. A custom-designed data acquisition module (TEM, Tower Electronics Module) is located below the calorimeter. Tracker towers are covered by a segmented scintillator anti-coincidence system (ACD), consisting of panels of plastic scintillator read out by wave-shifting fibers and photo-multiplier tubes, used to reject cosmic charged particle background (electrons, protons, heavier nuclei).
The ACD is also covered by a thermal blanket and micrometeorite shield. An aluminum Grid supports the detector modules and the data acquisition system and computers, which are located below the CAL modules. The LAT detector is described in more detail in \cite{4}. 

Each of the Tracker tower modules is composed of a stack of 19 stiff, lightweight carbon composite panels called trays: one bottom tray, 17 mid trays, and one top tray. A tray panel is a composite structure: an aluminum honeycomb core is closed by two carbon-fiber face sheets, bonded to four carbon-carbon close-outs to form a sandwich structure for mounting payload. On the top and bottom surfaces of the tray two kapton flexible circuits are glued to supply the bias potential to the silicon microstrip detectors (SSDs). The silicon detectors are bonded on both sides of a panel, with the strips on the top layer parallel to those on the bottom. Each layer hosts 16 SSDs in an array of four ladders, each one consisting of four wafers glued head to head and bonded together to form a single detector assembly. An array of tungsten (W) converter foils is glued to the bottom surface of all but the three lowest trays, between the panel and the flexible circuits and detectors, to match the active area of each wafer. The first 12 layers of tungsten are each 2.7$\%$ radiation length (r.l.) in thickness, while the following four layers are 18\% r.l. thick to increase the photon detection efficiency in the GeV and above energy range. Only the top and bottom trays support silicon detectors on one side only. Each tracker module has 18 x, y tracking planes, consisting of 2 layers (x and y) of single-sided silicon strip detectors. The total thickness of the TKR on-axis is 1.5 r.l.

Each CAL module is composed by 96 CsI(Tl) crystals ($2.7~ cm \times 2.0~ cm \times 32.6~ cm$) supported by a carbon fiber composite structure and read out from both ends with silicon PIN photodiodes. The crystals are optically isolated from each other and are arranged horizontally in 8 layers of 12 crystals each.  The total on-axis depth of the CAL is 8.6 r. l. Each calorimeter module layer is aligned $90^{\circ}$ with respect to its neighbors, forming an x,y (hodoscopic) array.
  The level of segmentation is sufficient to allow spatial imaging of the shower and accurate reconstruction of its direction because each CsI crystal provides three spatial coordinates for the energy deposited within: two discrete coordinates from the physical location of the crystal in the array and the third, more precise, coordinate determined by measuring the light yield asymmetry at the ends of the crystal along its longitudinal dimension.

Each crystal end is equipped with two photodiodes: a large one with cross section $1.5~ cm^2$, and a smaller one with cross section $0.25~ cm^2$. Each photodiodes is readout with two front end electronic channels with different gain, to cover the large dynamic range of energy deposition in the crystal. The large photodiodes cover the range 2 MeV - 1.6 GeV, while the small photodiodes cover the range 100 MeV - 100 GeV. A calibration charge injection signal can be fed directly to the front end of the pre-amplifiers \cite{onorbitcal}. The longitudinal segmentation enables energy measurements. From the longitudinal shower profile, we derive an unbiased estimate of the initial electron energy by fitting the measurements to an analytical description of the energy-dependent mean longitudinal profile \cite{simul}.

\section{Event selection and data reduction}
The LAT trigger system takes primarily inputs from the ACD, TKR, and CAL detectors. In addition, a periodic trigger is also configurated to perform the charge injection calibration. After an event passes the hardware trigger it is inspected by on-board software filters, each configured to identify events likely to be useful for one or more scientific or calibration purposes. There are 4 types of filters: 
\begin{itemize}
 \item Gamma: the purpose is to select gamma-ray candidates and events that deposit at least 20 GeV in the CAL;
 \item Heavy ion: the purpose is to perform calibration on high-energy scales;
 \item MIP: the purpose is to select non iteracting charged particles (protons);
 \item Diagnostic: the purpose is to select an unbiased event sample for filter and background performance studies
 \end{itemize}
The LAT on-board filters can be configured. If any filter accepts an event, it is included in the LAT data stream and forwarded to the solid state recording for transmission to the ground. The LAT event data contains TKR, CAL and ACD sub-system information, for istance TKR hit strips and Time Over Threshold (TOT), CAL and ACD amplitude signals, and trigger primitive.
 
The nominal science operation on orbit includes the Gamma, Diagnostic and Heavy ion filters. In particular, since the Gamma filter is configured to accept all events that deposit at least 20 GeV in the calorimeter, high-energy events, including electrons, are available for analysis on the ground.

To identify the CR electron component, we first used the ACD signal to select charged particle events that penetrated the detector. As for the analysis developed for extracting LAT photon data \cite{4}, the electrons selection essentially relies on the LAT capability to discriminate electromagnetic (EM) and hadronic showers based on their longitudinal and lateral development, as measured by both the TKR and CAL detectors. EM showers start developing in the TKR, while most of the energy is absorbed in the CAL. The measurement of the lateral shower development is a powerful discriminator between more compact EM showers and wider hadronic showers. 

The cuts have been tuned using a MC data sample of pure electrons and a realistic CR nuclei model \cite{simul}. We select on variables (measured or derived quantities) that map the distribution of TKR clusters around the main track, and in the CAL second-order moments of the energy distribution around the shower axis. A further selection derives from the different distributions of energy and hits in the ACD between EM and hadron-initiated showers. 

Similarly to the LAT photon background rejection analysis, the remaining necessary boost in the rejection
power is obtained by combining two probability variables that result from training classification trees (CT) to distinguish between EM and hadron events \cite{4}. Two CTs are used, one built
with TKR variables, and a second one based on CAL variables, which describe the complete event topology.
The variables given most weight by the CTs are the same or equivalent to those described above. The classifiers allow selection of the electrons through a multitude of parallel paths, each with different selections, that map the many different topologies of the signal events into a single, continuous probability variable that is used to simultaneously handle all valid selections. The TKR
and CAL electron probabilities are finally combined to create an energy-dependent selection that identifies electrons with greater efficiency and optimized background rejection with respect to a single sequence of cuts. 
 
The cuts used in the analysis are:
\begin{enumerate}
 \item[-] $cut_{qual}$: GammaFilter + at least 20 GeV energy deposit in the CAL + Trigger coincidence from ACD, TKR and CAL + at least one good track reconstructed in the TKR with angle with to respect the LAT Z-axis less than 72 deg + at least 7 r. l. in the CAL + reconstucted angle with respect to local zenith less than 105 deg to reject eventual Earth albedo photons + at least 1 ACD tile fired
 \item[-] $cut_{ACD}$: $cut_{qual}$ + energy deposit on ACD tiles equivalent to a single charged particle 
 \item[-] $cut_{TKR}$: $cut_{ACD}$ + large number of TKR clusters in a region around the good track and the average TKR Time Over Threshold (TOT) larger than the single charged particle value 
 \item[-] $cut_{CAL}$: $cut_{TKR}$ + narrow energy distribution in the CAL CsI crystals around the shower 
 \item[-] $cut_{CT}$: $cut_{CAL}$ + CT analysis 
\end{enumerate}

\begin{figure}[!hb]
  \centering
  \includegraphics[width=2.5in]{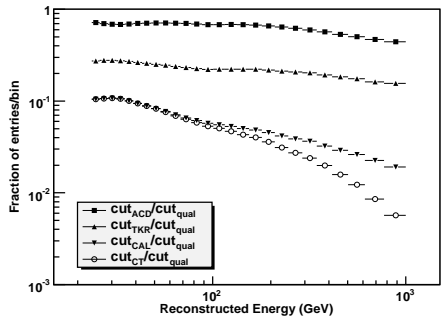}
  \caption{Fraction of events as function of reconstructed energy and cut types with respect to the $cut_{qual}$ (see text).}
  \label{fig1}
 \end{figure}
The data sample presented here has been collected from Aug. 4th, 2008 to Jan. 31st, 2009 in sky survey scanning operational mode. Fig.~\ref{fig1} shows the fraction of events as function of the reconstructed energy and cuts with respect to the $cut_{qual}$. The $cut_{ACD}$, $cut_{TKR}$ and $cut_{CAL}$ remove a large number of events in all energy regions, while the $cut_{CT}$ improves the electron/hadron separation above 100 GeV \cite{simul}. It is worth to point out that event selection so far is not optimized versus the incident angle of incoming particles, even though it is explicitly energy-dependent to suppress the larger high-energy background.
 
 \begin{figure}[!ht]
  \centering
  \includegraphics[width=2.5in]{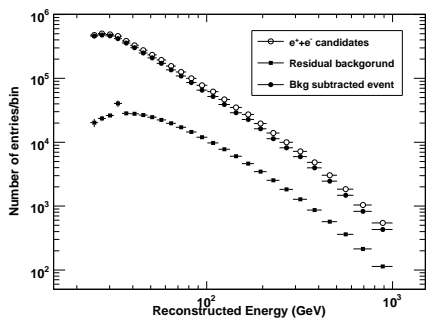}
  \caption{Number of events as function of energy. Open circles: candidate electrons+positrons; Filled squares: hadron contamination; Filled circles: background subtracted distribution.}
  \label{fig2}
 \end{figure}
The residual hadronic background in the final data sample has been estimated by means of a MC full simulation folding cosmic ray (background model) that the LAT encounters in space. The background model includes primary charged cosmic rays and Earth albedo photons \cite{4}. 
The residual hadronic background has been estimated from the average rate of hadrons that survive
electron selection in the simulations. Fig~\ref{fig2} shows the energy distribution for final candidate electrons (open circles) and the background (hadron) contamination (filled squares). The hadron contamination ranges from about 4\% at 20 GeV to 21\% at 1 TeV. More than 4M candidate electron events above 20 GeV have been selected, and 544 events before background subtraction have been collected in the last energy bin (770 GeV-1TeV). The final electron data sample has been evaluated subtracting bin by bin the background events from the candidate electrons (filled circles in Fig.~\ref{fig2}).  

Events considered for the electron analysis require to fail the ACD vetoes developed to select photon
events \cite{4}. This removes the vast majority of the potential gamma-ray contamination. In fact, the geometry factor for photons, determined with electron selection cuts, is less than 8\% of that for electrons at 1 TeV. The ratio between photon and electron fluxes is negligible at low energies and rises to around 20\% at the highest energy. This estimate is obtained from a simple, conservative extrapolation to high energies of the EGRET all-sky spectrum at GeV energies using a $E^{−2.1}$ power law \cite{egret}. The overall gamma contamination in the final electron sample is therefore always less than 2\%. This contamination is not subtracted from the final electrons data sample.

\section{Spectrum energy reconstruction}
\begin{figure}[!ht]
  \centering
  \includegraphics[width=2.5in]{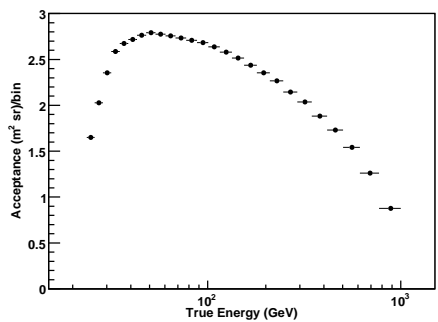}
  \caption{Electron Acceptance as function of true energy.}
  \label{fig3}
 \end{figure}

The electron/positron flux is evaluated by correcting the observed number of events with the instrument response function (IRF). The electron IRF (also called smearing matrix) is evaluated from Monte Carlo simulation, in which a trial spectrum of input pure electrons is simulated. The smearing matrix describes the full migration from the ``true'' energy bins (MC Energy) to the reconstructed ones. 
The elements of the smearing matrix represent the probabilities to observe a given effect that falls in a reconstructed energy bin from a cause in a given true energy bin. The smearing matrix also describes the losses of events due to cuts applied to extract the signal, which occur due to finite acceptance  (geometric factor) of the detector. The electron efficiency factor is computed as the ratio between the reconstructed number of events after applying the cuts and the number of generated events as function of the true energy. Fig.~\ref{fig3} shows the electron acceptance factor as function of true energy evaluated multiplying the efficiency for the area
and solid angle used for the generation of the Monte Carlo data sample.

In absence of energy dispersion, i.e. the true and the reconstructed energy are the same, the flux could be evaluated simply by dividing the number of observed counts by the detector acceptance. On the other hand, to include the finite energy dispersion effect the flux is evaluated using a deconvolution (unfolding) method. The deconvolution method adopted in the present analysis is based on Bayes' theorem \cite{dago}. The unfolding analysis allows to reconstruct the number of events belonging to each bin of true
energy. The integral flux in the $i-th$ energy bin (in units of events per unit area, solid angle and unit time) is then evaluated dividing the reconstructed events in the bin for the area of the sphere used for
the generation of the Monte Carlo data sample, solid angle and for the live time. The
differential flux (in units of events per unit area, solid angle, unit time and unit energy) is then evaluated dividing the integral flux for the corresponding energy bin width.

\begin{figure}[!ht]
  \centering
  \includegraphics[width=2.5in]{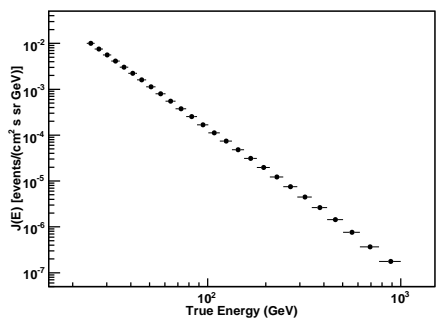}
  \caption{Electron and positron flux as function of energy. Only statistic errors are shown.}
  \label{fig4}
 \end{figure}

Fig.~\ref{fig4} shows the differential electron/positron flux as function of energy in the range from 24 GeV to 1 TeV. The overall LAT-Fermi electron spectrum can be fitted with a simple power law model, $E^{-\Gamma}$, with spectral index $\Gamma=3.04$. 

\begin{figure}[!ht]
  \centering
  \includegraphics[width=2.5in]{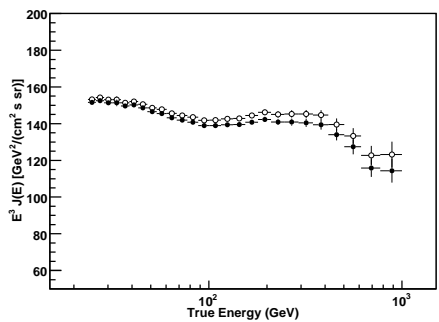}
  \caption{Electron and positron flux mulitplied by $E^3$ as function of energy. Open circles: average  energy in each is taken for $E^3$; Filled circles: median energy in each bin assumig a power law distribution with spectral index $\Gamma=3$ is taken for $E^3$. Only statistic errors are shown.}
  \label{fig5}
 \end{figure}
 
Fig.~\ref{fig5} shows the differential energy spectrum multiplied by $E^3$ in order to display eventual features that are otherwise difficult to discern in the Fig.~\ref{fig4}. Two kind of energy values have been considered for calculating $E^3$: the mean energy in each bin (open circles); and median energy in each bin assumig a power law distribution with spectral index $\Gamma=3$ (filled circles), since the overall energy spectrum follows a simple power law model with spectral index close to 3. The relative differences of the mean bin center with respect to the median to evaluate $E^3~J(E)$ is less than 8\% at 1 TeV.

The systematic uncertanties assessment provides a maximum error less than 20\% at 1 TeV~\cite{prl} . The absolute LAT energy scale, at this early stage of the mission, is determined with an uncertainty of $^{+5\%}_{-10\%}$. The systematic uncertanties are not shown in the energy spectrum.

\section{Conclusions}
The primary cosmic-ray electron/positron energy spectrum has been measured from 20 GeV to 1 TeV using six months of LAT-Fermi instrument. A detailed comparison of this result with the ones reported of many groups over the years is presented in Ref.~\cite{luca}. An interpretation of the results is presented in Ref.~\cite{dario}.


\end{document}